\documentclass[twocolumn,showpacs,aps,prl]{revtex4}
\usepackage{graphicx}
[1999/03/29 PSNFSS v.7.2
Times font as default roman
: S Rahtz]

\begin{document}
\renewcommand{\thefootnote}{\fnsymbol{footnote}}
\sloppy
\newcommand{\rp}{\right)}
\newcommand{\lp}{\left(}
\newcommand \be  {\begin{equation}}
\newcommand \ba {\begin{eqnarray}}
\newcommand \ee  {\end{equation}}
\newcommand \ea {\end{eqnarray}}

\title{Creep failures in heterogeneous materials}
\thispagestyle{empty}

\author{H. Nechad$^1$, A. Helmstetter$^2$, R. El Guerjouma$^1$ and D. Sornette$^{2,3,4}$}
\affiliation{$^1$  Groupe d'Etude de M\'etallurgie Physique et de Physique des Mat\'eriaux,
CNRS UMR5510 and INSA de Lyon, 20 Avenue Albert Einstein, 69621 Villeurbanne Cedex, France}
\affiliation{$^2$ Institute of Geophysics and Planetary Physics, 
University of California, Los Angeles, CA 90095}
\affiliation{$^3$ Department of Earth and Space Science,
University of California, Los Angeles, CA 90095}
\affiliation{$^4$ Laboratoire de Physique de la Mati\`{e}re Condens\'{e}e
CNRS UMR6622 and Universit\'{e} de Nice-Sophia Antipolis, 
B.P. 71, Parc Valrose, 06108 Nice Cedex 2, France}

\date{\today}

\begin{abstract}

We present creep experiments on fiber composite materials with 
controlled heterogeneity. 
Recorded strain rates and acoustic emission rates 
exhibit a power law relaxation in the primary creep regime (Andrade
law) followed by a power law acceleration up to rupture over up to four decades in time.
We discover that the failure time is proportional to the
duration of the primary creep regime, showing the interplay between the two regimes
and offering a method of rupture prediction. These experimental 
results are rationalized by a mean-field model of representative elements
with nonlinear visco-elastic rheology and with a large heterogeneity of strengths.
\end{abstract}


\maketitle

The damage and fracture of materials are technologically of enormous
interest due to their economic and human cost. Failure of composite systems 
is particularly important in naval, aeronautics and space industry. 
Despite considerable experimental \cite{L84,Aetal95,guarinocreep} and
theoretical work \cite{hr90,FM99,Z99} on fracture,
many questions have not been answered yet. 
Recently, statistical physicists have shown the existence of a power
law acceleration of acoustic emissions announcing the global failure of 
heterogeneous materials \cite{Aetal95,guarinocreep}, similar to the critical 
behavior of out-of-equilibrium phase transition \cite{Z99}, offering a way
to predict material failure \cite{Aetal95}.
 
This letter presents creep experiments on composite materials, which 
we explain using a simple model of representative elements, in the framework 
of fiber bundles models. Creep is the progressive deformation of a material
under constant load at a given temperature. 
Three creep regimes are usually observed. During the primary regime,
the strain rate decays as a power law with time following the application of the
stress (Andrade law) \cite{Andrade}. The secondary regime describes
a quasi-constant deformation rate, which evolves towards the
tertiary creep regime, if the stress and the temperature are high enough,
during which the strain rate accelerates up to rupture. 

The experiments are carried out on cross ply
glass/polyester composite materials and on Sheet Molding Compound
(SMC) composites. Two type of cross angle ply laminates are fabricated, 
denoted $[\pm62^o]$ and $[90^o/35^o]$, where the angles measure the directions 
of the glass fibers with respect to the loading direction, with a
fibre volume fraction of 75\%.
The SMC composites consist of a combination of polyester resin, calcium carbonate filler,
thermoplastic additive and random oriented short glass fibers, in the form of a sheet.
The relatively low fibre volume fraction, about 30\%, and the uncontrolled 
filler and reinforcement distribution during processing lead to a more 
heterogeneous structure for the SMC compared to the cross ply composites. 
The $[\pm62^o]$ and $[90^o/35^o]$ specimens have dimensions $14 \times 100 \times 2$ mm$^3$. 
The SMC samples are in the form of 120 mm barbell with 3 mm tickness. 
All specimens are subjected to a constant stress $s$ and temperature $T$ (below 
the glass transition of the matrix), which were fixed to $s=15$ MPa and 
$T=60^o$C for the $[\pm62^o]$ specimens,  $s=22$ MPa and $T=60^o$C for the 
$[90^o/35^o]$ specimens, and  $s =48$ MPa and $T=100^o$C for the SMC.
The creep tensile tests were performed using a servo-hydraulic mechanical testing system.
Constant tensile load was applied and the resulting strain and acoustic emissions
were recorded until final rupture. Acoustic emissions (AE) is a standard technique 
to monitor the evolution of damage in composites, due to matrix cracks, fiber matrix 
debounding, fiber breaks, and delaminations \cite{EG01}.
We used a Mistras data acquisition system by Physical Acoustics 
Corporation with 2 resonant sensors (200 kHz - 1 MHz).

Normal primary creep transients followed by secondary and tertiary creep were 
observed for almost all samples, both for the strain rate (Fig. \ref{dedt62}) 
and for the AE rate (Fig. \ref{EACR62n4}).
The decrease of the strain rate and EA rate
in the primary creep regime can be described by Andrade's law \cite{Andrade}
$de/dt \sim t^{-p}$, with an exponent $p$ in the range 0.2-1.4 for the 15
samples tested \cite{vl}. 
The crossover for small times is probably due to the fact that the stress
progressively increases up to about 10 sec after the start of the experiment.
A quasi-constant strain rate (steady-state or secondary creep) is 
observed over an important part of the total creep time, followed by
an increase of the creep rate up to failure in the tertiary creep regime.
Creep strains at fracture are around 40\% for angle 
cross ply composites and $\approx 4\%$ for the SMC.
The acceleration of the strain rate before failure is well fitted by a
power-law singularity $de/dt \sim (t_c-t)^{-p'}$ 
with $p'$ in the range 0.3-1.1 depending on the sample \cite{vl}.
The critical time $t_c$ determined from the fit of the data
with a power-law is close to the observed failure time.
Our experiments confirm over large time scales covering up to four orders
of magnitude in time
previous announcement of power laws in the tertiary creep regime, which
were established over more limited time scales \cite{guarinocreep}. 
We also obtain the same temporal evolution for the AE energy rate, 
with larger fluctuations for the energy rate than for the event rate 
due to the existence of a power-law distribution of AE energies.
We found no qualitative differences between the behavior of the 
 $[\pm62^o]$ and $[90^o/35^o]$ cross ply composites.
The values of $p$ and $p'$ are on average a little larger for the SMC 
than for the cross ply composites, possibly due to the larger
heterogeneity of the SMC or to the different values of the applied stress.

There is a huge variability of the failure time from one sample to another one, 
for the same applied stress, as shown in Fig. \ref{tmtc}.
This figure shows that the transition time $t_m$ between the primary creep 
regime and the tertiary regime, measured by the minimum of the strain rate,
is proportional to the rupture time $t_m \approx 2/3~t_r$.
We also found a negative  correlation between the Andrade exponent $p$ 
and the rupture time $t_r$ \cite{vl}. 
These observations shows that damage in the primary regime 
impacts on its subsequence evolution in the secondary and tertiary regimes.
This suggests a way to predict the failure time from the observation of the strain 
rate or AE rate during the primary and secondary creep regimes, before the acceleration of 
the damage leading to rupture.

Creep observations have been modelled in terms of visco-elastic fibers, with deterministic 
dynamics and quenched disorder \cite{Kun}. This model reproduces a power law singularity 
of the strain rate before failure with $p'=1/2$ in the case of a uniform distribution of 
strengths \cite{Kun} but does not explain Andrade's law for the primary creep. 
Here, we start from the model of \cite{Kun} and 
enrich it with a more realistic rheology and heterogeneity,
in order to account simultaneously for Andrade's law in the primary creep and 
for the power-law singularity of the strain rate before failure.
We view a composite system as made of a large set of representative elements (RE),
each element comprising many fibers with their interstitial matrix. 
The applied load is shared democratically between all RE. This 
assumption has been shown to be a good approximation 
of the elastic load sharing for sufficiently
heterogeneous materials \cite{rouxhild02}.
Each RE is modelled as a non-linear Eyring dashpot \cite{LiuRoss} in parallel 
with  a linear spring of stiffness $E$. 
The Eyring rheology, which is standard for fiber
composites, consists at the microscopic level
in adapting to the matrix rheology the theory of reaction rates
describing processes activated by crossing potential barriers.
A given RE fails when its elongation/deformation 
$e$ reaches a threshold. The rupture thresholds are distributed according to the cumulative 
distribution $P(e)$ given by
$P(e)=1-\left(e_{01} / (e +e_{02}) \right)^{\mu}$,
where $e_{01}$ and $e_{02}$ are two constants with $e_{01} \leq
e_{02}$. The fraction $1-(e_{01}/e_{02})^{\mu}$ breaks as
soon as the stress is applied.
The power-law distribution $P(e)$ for large $e$ is  
motivated by the large distribution of failure times for the same applied stress
(Fig. \ref{tmtc}). The exponent $\mu>1$ controls the amplitude of the 
frozen heterogeneity of the RE strengths.

The equation controlling the deformation $e(t)$ of each surviving
RE is 
\be
{de \over dt} = K \sinh \left({\beta s \over 1-P(e)} -\beta Ee \right) 
\label{eyringdash} 
\ee
with the initial condition $e(t=0)=0$. The fraction of unbroken RE is $1-P(e)$ 
and $s/( 1-P(e))$ is the stress applied on each unbroken RE.

The system defined by (\ref{eyringdash}) is stable (no global rupture) if the differential 
equation (\ref{eyringdash}) has a stationary solution $de/dt=0$ with $e>0$, i.e., 
if the equation $\left((e+e_{02})/e_{01}\right)^{\mu} = (E / s) e$
has a non-trivial solution. This defines a threshold $s^*$ below which the
strain converges asymptotically to a constant and above which 
$de/dt$ grows without up to rupture.

In the primary regime  $e \ll e_{02}$ thus 
$(e+e_{02})^{\mu} \approx {e_{02}}^{\mu} (1 + \mu e /e_{02})$.
If the stress on the dashpot is small, we can replace $\sinh$ by $\exp/2$. 
With these approximations, the differential equation
({\ref{eyringdash}) has the solution 
\be
{de \over dt} = {K  \over 
2e^{-\beta s \left({e_{02} \over e_{01}} \right)^\mu}
+ t K \beta \left(
E - {\mu s \over e_{02}}   \left({e_{02} \over e_{01}} \right)^\mu \right) }~.
\label{pc}
\ee
Expression (\ref{pc}) predicts that, if the stress is not too large, 
${de/ dt}$ is of the Andrade form $\sim t^{-p}$, with an exponent $p=1$ at early times.
For larger $s$, the strain rate starts to accelerate as soon as the load is applied.
Note that the observation of Andrade's power-law creep in this model 
does not involve any failure of RE and is thus independent of
the choice of the distribution of rupture thresholds $P(e)$.

In the tertiary creep regime, we can neglect $e_{02}$ compared with $e$. 
Close to failure, for large $e$, the linear term $E e$ is small compared 
with $s/(1-P(e))={ s \over {e_{01}}^\mu} (e+e_{02})^{\mu}$ if $\mu>1$.
This leads to the equation
\be
{de \over dt} \approx {K \over 2} \exp \left( {\beta s {e}^{\mu} \over {e_{01}}^\mu} \right)~.
\label{dif1ss}
\ee
Its solution is, to leading logarithmic order,
\be
{d e \over dt} = {A \over \mu} ~\left[ -\ln(t_c-t)\right]^{{1 \over \mu}-1}~
{1 \over t_c -t}~, \label{mgmss}
\ee
where  $A=e_{01} \left(\beta s \right)^{-1/\mu}$.
Contrary to the primary regime, the heterogeneity of the rupture threshold
is an essential ingredient for the power-law singularity before failure,
but the leading power-law term with $p'=1$ in  (\ref{mgmss}) does not depend
on the exponent $\mu$ characterizing heterogeneity.
The acceleration toward failure is due to the positive feedback effect of broken RE,
which increases the stress and strain on the unbroken RE leading
to the global failure of the system.

Figure \ref{num1} shows the numerical solution of equation 
(\ref{eyringdash}) together with the approximate analytical solutions (\ref{pc}) 
in the primary creep and (\ref{mgmss}) close to failure,
for different values of the applied stress $s$.
In the primary creep regime close to the rupture threshold $s\approx s^*$,
we observe numerically an apparent exponent $p<1$, smaller than
predicted by  (\ref{pc}), which can explain the values of $p$
found experimentally \cite{vl}. For a stress $s \gg s^*$, the strain rate 
accelerates immediately when the load is applied. For $s<s^*$, the $p$-value  
decreases in the model between 1 and 0 as the applied stress increases. 
The duration of the primary creep also decreases with $s$.
The model thus explains the correlation found experimentally between the $p$-value and
the failure time \cite{vl}.

In the tertiary regime, for $s\gg s^*$, we find numerically that expression (\ref{dif1ss})
is a good approximation very close to failure $t\approx t_c$. 
But for $s\approx s^*$, there is a crossover further from failure 
with an apparent exponent $p'=0.9$. This simple model thus reproduces 
both power-laws in the primary and tertiary creep regimes, with an
apparent exponent $p \leq 1$ for the primary creep, 
and with $p'=1$ for the tertiary regime, except
for a crossover with an apparent exponent $p'$ a little smaller than 1. 
This crossover with $p'<1$ is however not sufficient to explain the observations 
of $p'=0.93$ over 4 orders of magnitude in time $t_c-t$ (Fig. \ref{dedt62}).  

The failure time has a power-law singularity $\sim
(s-s^*)^{-1/2}$ for $s\approx s^*$, as found previously \cite{Kun}
for the model with a linear dashpot,
and decays exponentially for $s\gg s^*$ \cite{vl}.
The transition time $t_m$ (minima of the strain rate) is equal to $t_c/2$.
This result recovers the proportionality of $t_m$ 
and $t_c$ found experimentally, but predicts a duration for the primary creep 
shorter than the observations $t_m \approx 2t_c/3$ (Fig. \ref{tmtc}).

In conclusion, we have shown that the interactions between the RE elements together
with a large heterogeneity and a simple nonlinear rheology is sufficient to
explain qualitatively and quasi-quantitatively our experiments. This model replaces the need
for complex memory effects (such as the integro-differential Schapery long-memory 
formalism \cite{cardon}) often invoked in the composite literature.
A natural improvement of the model would be to relax the democratic
load sharing rule as in \cite{Kun} in order to introduce realistic elastic interactions.
This improvement may provide a more realistic value of  $p'$  and of
the constant of proportionality between $t_m$ and $t_c$ (Fig. \ref{tmtc}).

We acknowledge useful discussions with F. Sidoroff and
A. Agbossou. This work is partially supported by
the James S. Mc Donnell Foundation 21st century scientist
award/studying complex system and by the french department of 
research under grant N$^o$ 207.

{}

\clearpage
\begin{figure}
\includegraphics[width=1\textwidth]{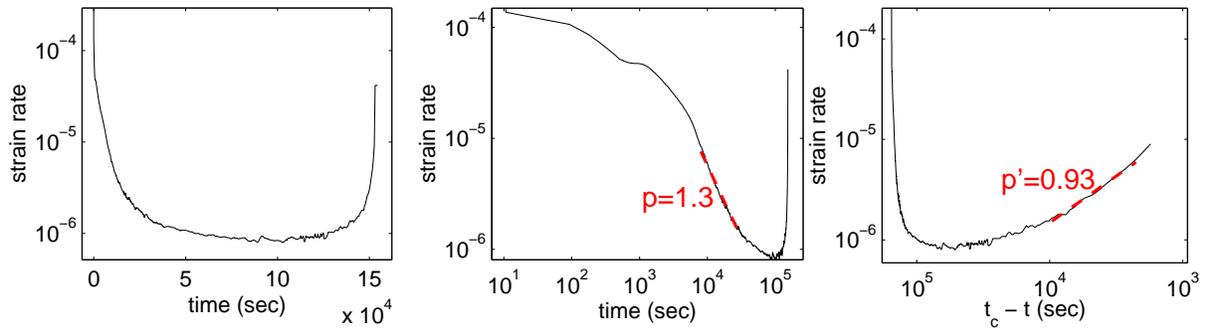}
\parbox{1\textwidth}{\caption{\label{dedt62} Creep strain rate for a $[\pm62^o]$ specimen.
(a) linear time scale, (b) logarithmic time scale to test for the 
existence of Andrade's law in the primary creep, (c)
logarithmic time scale in $t_c-t$ to test the time-to-failure power law 
 in the tertiary creep.}}
\end{figure}

\hskip -5cm

\begin{figure}
\includegraphics[width=1\textwidth]{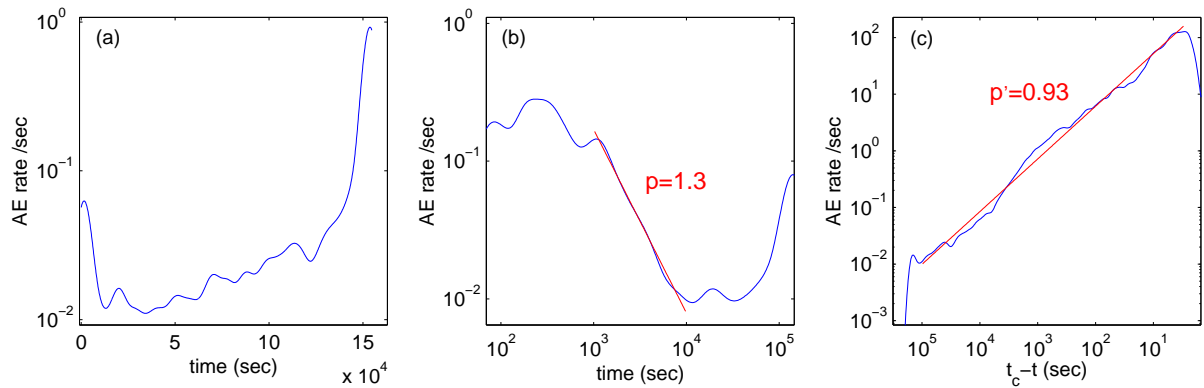}
\parbox{1\textwidth}{\caption{\label{EACR62n4} Rate of AE events for a $[\pm62^o]$ specimen.
The three panels are as in Fig. \ref{dedt62}.}}
\end{figure}

\clearpage
\begin{figure}
\includegraphics[width=0.5\textwidth]{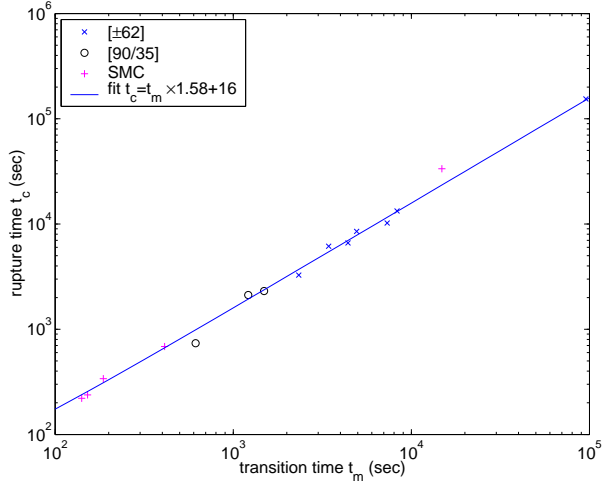}
\parbox{1\textwidth}{\caption{\label{tmtc} Relation between the time $t_m$ of the minima of  the 
strain rate and the rupture time $t_r$, for all samples.}} 
\end{figure}

\begin{figure}
\includegraphics[width=1\textwidth]{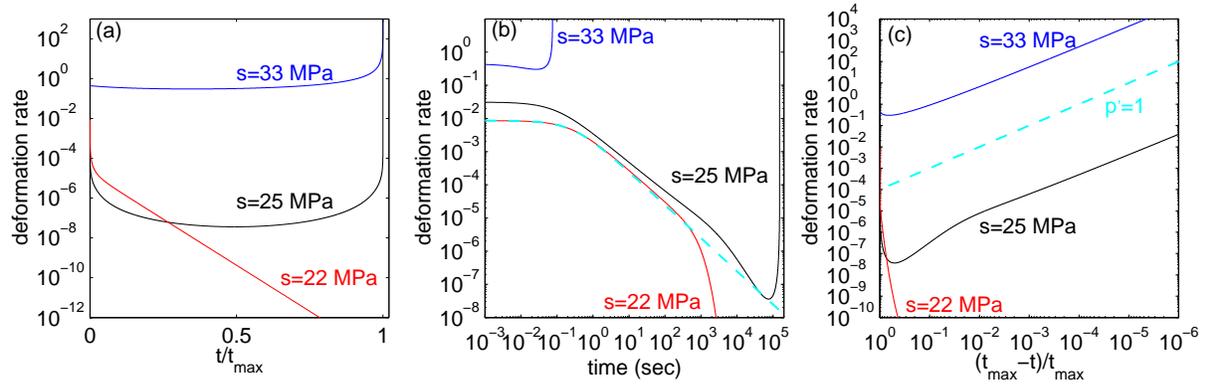}
\parbox{1\textwidth}{\caption{\label{num1} Strain rate $de/dt$ given by (\ref{eyringdash})  
for different values of the stress $s$, and with parameters $E=20$ GPa, 
$\mu=1.2$, $e_{01}=0.003$, $e_{02}=0.015$, $\beta=50$ GPa$^{-1}$ 
and $K=10^{-5}$ sec$^{-1}$. 
Panel (b) illustrates Andrade's law in the 
primary regime, with exponent $p \approx1$ for $s=22$ MPa 
and $p \approx 0.8$ for  $s=25$ MPa. The dashed line 
is the approximate solution (\ref{pc}) of (\ref{eyringdash}) with $s=22$ MPa.
Panel (c) shows the power law acceleration of $de/dt$
before failure for $s=25$ MPa and $s=33$ MPa, with $p'\approx 1$  asymptotically.
In (a) and (c) the time is normalized by the rupture time
for $s=25$ MPa and for $s=33$ MPa, and by the time when 
$de/dt$ decreases below $10^{-14}$ for  $s=22$ MPa.}}
\end{figure}

\end{document}